\documentclass[12pt,a4]{article}
\usepackage[full]{textcomp}
\usepackage{amsthm,amsmath,latexsym,amssymb,amsfonts,amscd}
\usepackage{graphics,lscape,fancyhdr,array,stmaryrd,euscript}
\usepackage{newtxtext,newtxmath}
\usepackage{verbatim}
\usepackage{array}
\usepackage[all]{xy}

\usepackage[numbers,sort&compress]{natbib}
\usepackage[nottoc]{tocbibind}

\usepackage{verbatim,slashed}
\numberwithin{equation}{section}
\usepackage{hyperref,setspace}

\usepackage{mathrsfs}

\newcommand{\pl}{\partial}

\newcommand{\be}{\begin{equation}}
\newcommand{\ee}{\end{equation}}
\newtheorem{lemma}{Lemma}

\newcommand{\bry}{{{\bar{y}}}}

\newcommand{\fud}[2]{{}^{#1}{}_{#2}\,}
\newcommand{\fdu}[2]{{}_{#1}{}^{#2}\,}

\newcommand{\besubeqs}{\begin{subequations}}
\newcommand{\esubeqs}{\end{subequations}}

\begin{document}
\pagenumbering{gobble}
\hfill
\vskip 0.01\textheight

\begin{center}
{\Large\bfseries 
Flat Self-dual Gravity}

\vskip 0.03\textheight
\renewcommand{\thefootnote}{\fnsymbol{footnote}}
Kirill \textsc{Krasnov}${}^{a}$ \&  Evgeny \textsc{Skvortsov}\footnote{Research Associate of the Fund for Scientific Research -- FNRS, Belgium}${}^{b,c}$ 
\renewcommand{\thefootnote}{\arabic{footnote}}
\vskip 0.03\textheight

{\em ${}^{a}$School of Mathematical Sciences, \\University of Nottingham, NG7 2RD, UK}\\
\vspace*{5pt}
{\em ${}^{b}$ Service de Physique de l'Univers, Champs et Gravitation, \\ Universit\'e de Mons, 20 place du Parc, 7000 Mons, 
Belgium}\\
\vspace*{5pt}
{\em ${}^{c}$ Lebedev Institute of Physics, \\
Leninsky ave. 53, 119991 Moscow, Russia}\\

\vskip 0.02\textheight

\begin{abstract}
We construct a new covariant action for "flat" self-dual gravity in four spacetime dimensions. The action has just one term, but when expanded around an appropriate background gives rise to a kinetic term and a cubic interaction. Upon imposing the light-cone gauge, the action reproduces the expected chiral interaction of Siegel. The new action is in many ways analogous to the known covariant action for self-dual Yang-Mills theory. There is also a sense in which the new self-dual gravity action exhibits the double copy of self-dual Yang-Mills structure.
\end{abstract}
\end{center}

\newpage
\tableofcontents
\newpage
\section{Introduction}
\pagenumbering{arabic}
\setcounter{page}{1}
Self-dual Yang-Mills (SDYM) and self-dual gravity (SDGR) have many remarkable properties. Both can be viewed as truncations of the YM and GR that keep only a subset of the solutions (and also interactions) of the full theories. In the case of SDYM these are "instantons", which are gauge field configurations whose field strength is purely anti-self-dual, so that the self-dual part of the curvature 2-form vanishes
\be\label{intro-sdym-feqs}
F^a_{SD}=0\,,
\ee
Here $a$ is the Lie algebra index, and the form indices are suppressed. Gauge fields satisfying this first order differential equation are then also automatically solutions of the second-order YM field equations $d_A^\mu F^a_{\mu\nu}=0$, where $d_A$ is the covariant derivative with respect to the connection $A$. In the case of SDGR the solutions are Einstein (i.e. having the vanishing tracefree part of the Ricci tensor) metrics with purely anti-self-dual Weyl curvature. While this appears to be a second-order in derivatives condition on the metric, there exists reformulations in which the gravitational self-duality equations are first order in derivatives, see below. One can then see that, similar to the case of SDYM, solutions of the first-order self-duality equations are also automatically solutions of the second-order field equations of full GR. 

Both SDYM and SDGR can also be studied as quantum theories. They are both one-loop exact and quantum finite, see \cite{Krasnov:2016emc} for a discussion of this point. Both theories can be usefully characterised (and contrasted with their full YM and GR cousins) in terms of the scattering amplitudes that they produce. At tree level, the only non-vanishing (for complex momenta) amplitude is at 3 particles. This amplitude is chiral. In our conventions this is the $--+$ amplitude. All higher point amplitudes vanish at tree level. At one-loop level, all same (negative) helicity amplitudes are non-vanishing. We refer the reader to e.g. \cite{Krasnov:2016emc} and references therein for more information.

There exist several known formulations of SDYM and SDGR. In former case, what appears to be the most useful covariant formulation is one first proposed by Chalmers and Siegel \cite{Chalmers:1996rq}. The action can be written as
\be\label{intro-sdym}
S_{SDYM}= \int \Psi^{ia} H^i \wedge F^{a}.
\ee
Here $H^i, i=1,2,3$ is a triple of self-dual two-forms that are also known as 't Hooft symbols. These objects are dimensionless and satisfy the algebra of quaternions
\be
H^i_{\mu}{}^\alpha H^j_\alpha{}^\nu = - \delta^{ij} \delta_\mu{}^\nu + \epsilon^{ijk} H^k_\mu{}^\nu\,.
\ee
The field $\Psi^{ia}$ is a triple of Lie algebra valued scalars, and $F^a$ is the Lie algebra valued field strength 2-form. Note that the combination $\Psi^{ia} H^i$, for varying $\Psi^{ia}$, parametrises a generic Lie algebra valued self-dual 2-form. Varying the action with respect to $\Psi^{ia}$ one gets the equations $H^i \wedge F^a=0$, which imply (\ref{intro-sdym-feqs}). There are other, non-covariant formulations of SDYM, some of which also feature in \cite{Chalmers:1996rq}. See also \cite{Bittleston:2020hfv} for a more recent discussion, and also on how to obtain (non-covariant) actions from the twistor space. 

There exists a covariant formulation of SDGR in flat space, somewhat hidden in the discussion in \cite{Siegel:1992wd}, and discussed much more explicitly in \cite{AbouZeid:2005dg}. We will review this action below. It is considerably different from the SDYM action (\ref{intro-sdym}). Another action for SDGR, applicable for describing anti-self-dual Einstein metrics with non-zero scalar curvature was studied in \cite{Krasnov:2016emc}. It is based on a "pure connection" description of gravitational instantons, and is closer to the SDYM Chalmers-Siegel action (\ref{intro-sdym}). The main purpose of this paper is to describe a covariant action for SDGR that is applicable for metrics with zero scalar curvature. Thus, the theory that is the main object of this paper can be referred to as "flat" SDGR. The action we describe is new, but is close to that in \cite{Krasnov:2016emc}, and can be motivated from the latter by a contraction procedure that drops the connection-connection term from the field strength. The new action is much closer to the SDYM action (\ref{intro-sdym}) than the previously available "flat" SDGR formalism \cite{Siegel:1992wd}, \cite{AbouZeid:2005dg}.

In its simplest form that is suitable for expanding around the flat space ${\mathbb R}^4$ the new action has a remarkably simple form 
\begin{align}\label{action-intro}
    S[a,\Psi]&= \tfrac12 \int \Psi_{ij}\, da^i \wedge  da^j\,.
\end{align}
Here $i,j=1,2,3$ and $\Psi^{ij}$ is a field that is symmetric and tracefree $\Psi^{ij}\delta_{ij}=0$. The action is a functional of exact two-forms $da^i$, and to obtain Euler-Lagrange equations one varies with respect to both $\Psi^{ij}$ and $a^i$. In the main text we will give a formulation that works also on arbitrary closed manifolds. In this case the exact two-forms in the action are replaced by arbitrary closed two-forms that are varied by exact forms. 

The background that describes the flat space and that gives a starting point for the perturbative expansion is
\be
da^i = M^2 H^i\,,
\ee
where $H_i$ are the already encountered self-dual 2-forms, and $M^2$ is the parameter of dimension mass two, introduced for dimensional reasons. As will be further discussed below, a triple of 2-forms $H^i$ (that is suitably non-degenerate) uniquely determines a conformal class of a metric for which $H^i$ are self-dual. Then, the volume form is fixed as a multiple of $\delta_{ij} H^i \wedge H^j $. Therefore, $H^i$ determines a unique metric, and this is how the fields contained in the action (\ref{action-intro}) determine the metric. The action (\ref{action-intro}) expanded around this background then reads
\begin{align}\label{action-pert-intro}
   S_{\text{SDGR}}&= \int \Psi_{ij} \left(M^2 H^i\wedge  da^j +\tfrac12  da^i \wedge  da^j\right)\,.
\end{align}
The first term can be shown, see below, to give the correct action for free gravitons, and the second term is a simple cubic vertex. It is instructive to compare this to the SDYM action (\ref{intro-sdym}) that we write in the form that exhibits strongest similarity to (\ref{action-pert-intro})
\begin{align}\label{action-SDYM}
   S_{\text{SDYM}}&= \int \Psi_{ia} \left(H^i\wedge  dA^a +\tfrac12  f\fud{a}{bc}H^i\wedge A^b \wedge  A^c\right)\,.
\end{align}
Here $A^a$ is a one-form connection taking values in a Lie algebra with structure constants $f\fud{a}{bc}$. The new SDGR action thus exhibits the double copy structure \cite{Bern:2010ue} of gravity rather explicitly. Indeed, this is particularly prominent in the first, kinetic terms of both actions. It is clear that one passes from SDYM to SDGR by replacing the Lie algebra index in $\Psi^{ia}, A^a$ by the "gravitational" index $i$. This is precisely what the double copy procedure is supposed to do, replacing "colour" by "kinematics". It is this similarity between the SDYM and the new SDGR action that is, in our opinion, one of the most attractive features of the new formulation. In the main text we will see that the similarity between the two theories extends further and SDYM and SDGR in the form (\ref{action-intro} are similar in many aspects. 

We will also show that in the light-cone gauge the new action reduces to the well-known Siegel action \cite{Siegel:1992wd}
\begin{align}
    \label{action-LC-intro}
    S&= \int (\Phi_{-2}\square \Phi_{+2} + M_{Pl}\,\epsilon_{A'B'}\epsilon_{C'D'} \Phi_{+2}\pl^{+A'}\pl^{+C'}\Phi_{-2}\pl^{+B'}\pl^{+D'}\Phi_{-2})\,,
\end{align}
where $\Phi_{\pm2}$ are two scalars representing helicity $\pm2$ states of the graviton. 

A part of our motivation in this paper stems from the study of Chiral Higher Spin Gravity \cite{Metsaev:1991mt,Metsaev:1991nb,Ponomarev:2016lrm,Skvortsov:2018jea,Skvortsov:2020wtf}, which is the minimal extension of gravity by massless higher spin fields. Its action contains the SDYM and SDGR actions that are coupled to higher spin fields featuring their own interactions too. The theory is UV-finite at one-loop despite the naive non-renormalizability of higher derivative higher spin interactions \cite{Skvortsov:2018jea,Skvortsov:2020wtf,Skvortsov:2020gpn} and captures a subset of correlation functions of Chern-Simons matter theories \cite{Skvortsov:2018uru} via AdS/CFT. The action (\ref{action-intro}) arose as a by product of \cite{Krasnov:2021nsq}, where covariant actions featuring gauge and gravitational interactions of higher spin fields were constructed. But these higher spin considerations will not play any role in this paper. We motivate and introduce the action (\ref{action-intro}) staying firmly in the context of usual gravity and its self-dual truncation. 

The outline of the paper is as follows. In section \ref{sec:sd-graflat} we present the new formulation of the SDGR in flat space and discuss its relation to mathematical results and to other formulations. In section \ref{sec:perturb} we discuss the perturbative aspects such as gauge-fixing, amplitudes and the light-cone gauge. Finally, in section \ref{sec:algebra} we identify the gauge algebra behind flat SDGR, which turns out to be a certain contraction of $so(3,2)$ that is different from the Poincare algebra. 

\section{Self-Dual Gravity}
\label{sec:sd-graflat}

The discussion of this section is phrased in terms of Euclidean signature gravity. Indeed, anti-self-dual Einstein metrics are only non-trivial (i.e. not maximally symmetric) when the signature is Euclidean or split. Alternatively, one can interpret the constructions below as those for complexified GR. 

\subsection{Hyper-Kähler manifolds}
It is well-known that a zero scalar curvature four-dimensional gravitational instanton is a hyper-Kähler manifold. We recall that a hyper-Kähler manifold of dimension $4k$ is an Einstein manifold that is Kähler with respect to 3 different complex structures $I,J,K$ that anti-commute and form the algebra of imaginary quaternions $IJ=K$. This means that the manifold has a quaternionic structure, which in turn implies that the holonomy is contained in ${\rm Sp}(k)$ and the manifold has zero scalar curvature. In the case $k=1$ we have ${\rm Sp}(1)={\rm SU}(2)$ and a hyper-Kähler manifold is a complex 2-dimensional Calabi-Yau manifold. Any compact such manifold is either a $K3$ surface or a torus $T^4$.

Given that the holonomy of a 4-dimensional hyper-Kähler manifold is contained in one of the two chiral halves ${\rm SU}(2)$ of the 4-dimensional rotation group ${\rm SO}(4)$, it is always possible to choose a gauge in which the chiral half of the spin connection for the other ${\rm SU}(2)$ vanishes. This gives a very convenient starting point for describing such manifolds. In the mathematical literature such a description is well known. It has been sketched in \cite{Donaldson2006TwoformsOF} and described in details in \cite{FINE_2017}.

We start by describing the result that appears as Lemma 3.1 in \cite{FINE_2017}.
\begin{lemma} Let $X$ be a 4-manifold and $(\omega^1,\omega^2,\omega^3)$ a triple of closed 2-forms on $X$ that satisfy 
\be \label{simplicity}
\omega^i\wedge \omega^j = 2\delta^{ij} \mu\,,
\ee
where $\mu$ is some non-vanishing 4-form on $X$. Then $X$ carries a hyper-Kähler metric, which is characterised by the fact that all $\omega^i$ are self-dual and the volume form is $\mu$.
\end{lemma}

The proof is based on several steps, see \cite{FINE_2017}, and we just sketch the main points. Given a triple of non-degenerate 2-forms $\omega^i$ on $X$ there exists a unique conformal class of a metric on $X$ which makes $\omega^i$ self-dual. If the wedge product on the subbundle in $\Lambda^2$ spanned by $\omega^i$ is definite, then the conformal metric one obtains is of a Riemannian signature. Choosing a volume form one then gets a metric. In the case of 2-forms satisfying (\ref{simplicity}) the natural volume form to complete the definition of the metric is $\mu$. Overall, with (\ref{simplicity}) satisfied we get a Riemannian signature metric $g_\omega$ defined by $\omega^i$.

The second important statement is that when the 2-forms are closed $d\omega^i=0$, and satisfy (\ref{simplicity}), the self-dual part of the spin connection vanishes. This immediately implies that $g_\omega$ is Einstein of zero scalar curvature, and with only a half of its Weyl curvature possibly non-zero. Indeed, the fact that the self-dual part of the spin connection is zero implies that also its curvature 2-form is zero. However, as is well-known, see e.g. \cite{Krasnov:1970bpz}, Chapter 5, the curvature of only one of the two chiral halves of the spin connection contains enough information to impose the Einstein condition. Thus, in general, the curvature of the 
self-dual part of the spin connection can be decomposed into its self- and anti-self-dual parts (as a 2-form). The self-dual part then encodes one of the two chiral halves of the Weyl curvature, as well as the scalar curvature. The anti-self-dual part encodes the tracefree part of Ricci. When both these parts vanish we have a zero scalar curvature Einstein manifold, whose self-dual part of the Weyl curvature also vanishes. This is a gravitational instanton.

We can then propose a variational principle that leads to (\ref{simplicity}) as one of the Euler-Lagrange equations. The action we take is
\be\label{action}
S[\Psi,\omega] =\frac{1}{2} \int \Psi^{ij} \omega^i\wedge \omega^j\,.
\ee
Here $\Psi^{ij}$ is a field that is symmetric $\Psi^{ij}=\Psi^{(ij)}$, which is moreover assumed to be tracefree $\delta_{ij}\Psi^{ij}=0$. Varying the action with respect to this field produces the condition (\ref{simplicity}). 

The action (\ref{action}) can also be varied with respect to $\omega^i$. It should be kept in mind, however, that the 2-forms $\omega^i$ are not free to vary, as they are assumed to be closed. This means that it is natural to allow only a variation of each one of them by an {\bf exact} form. Thus, the definition of the action is completed by requiring that the variations in the space of 2-forms are by an exact form
\be
\delta \omega^i = d a^i\,.
\ee
This is similar to the variational principles considered by Hitchin in \cite{Hitchin:2001rw}. With this in mind, the Euler-Lagrange equation arising by varying $\omega^i$ are
\be
d \Psi^{ij} \wedge \omega^j=0\,,
\ee
where we used the fact that $\omega^i$ are closed. This equation describes a propagation of the field $\Psi^{ij}$ in the background of the gravitational instanton described by $\omega^i$.

The action (\ref{action}) is invariant under diffeomorphisms 
\begin{align}
    \delta \omega^i= \mathcal{L}_\eta \omega^i, \qquad \delta \Psi^{ij}=\mathcal{L}_\eta \Psi^{ij}\,,
\end{align}
where $\eta\equiv \eta^\mu$ and $\mathcal{L}_\eta =\{d, i_\eta\}$ is the Lie derivative. Also, if one decides to parametrise 2-forms $\omega^i$ in a given cohomology class, then the variation in each cohomology class is an exact form $da^i$. It is clear that the 1-forms $a^i$ are defined modulo exact forms $d\theta^i$, where $\theta^i$ are zero-forms. 
\subsection{Relation to SDGR on anti-de Sitter space.} 
There is another covariant action for SDGR \cite{Krasnov:2016emc}, which is better suited for expansion over (anti-) de Sitter space. The flat action we described above can be understood as arising by a "contraction" procedure from the action \cite{Krasnov:2016emc}.

Action \cite{Krasnov:2016emc} has a related field content and reads
\begin{align}\label{action-lambda}
    S[\Psi,A]&=\tfrac12\int \Psi^{ij}  F^i \wedge F^j\,.
\end{align}
The action is a functional of the "Lagrange multiplier" field $\Psi^{ij}$, which is the same field that appears in (\ref{action}), and an ${\rm SO}(3)$ connection $A^i$. The object $F^i$ is the curvature 2-form 
\begin{align}
    F^i&= d A^i+\frac{1}{2}\epsilon^{ijk} A^j\wedge A^k\,.
\end{align}
The symmetries of this theory are the local ${\rm SO}(3)$ symmetries and diffeomorphisms. 

The action (\ref{action-lambda}) describes anti-self-dual Einstein metrics with non-zero scalar curvature. The logic that leads to this conclusion is very similar to that described in the previous subsection. Thus, the equation one obtains by varying with respect to $\Psi^{ij}$ is 
\be\label{F-eqs}
F^i\wedge F^j = 2\delta^{ij}\mu\,.
    \ee
Similar to the flat case, in which the role of $F^i$ is played by the symplectic forms $\omega^i$, the triple $F^i$ of curvature 2-forms can be used to define a Riemannian signature metric $g_F$. It has the unique conformal class that makes the 2-forms $F^i$ self-dual, and it has the volume form $\mu$. The second step is also similar to one we had in the flat case. When (\ref{F-eqs}) are satisfied, the Bianchi identity $d_A F^i=0$ implies that the connection $A^i$ is the self-dual part of the spin connection for the metric $g_F$. However, unlike in the flat case, this connection no longer needs to vanish. Instead, its curvature, which coincides with the self-dual part of the Riemann curvature, is in an appropriate sense constant, which means that only the scalar curvature, and possibly the anti-self-dual part of the Weyl curvature is non-vanishing. This means that the metric $g_F$ is a gravitational instanton. For more details on this construction we refer the reader to \cite{Krasnov:2016emc} and also to \cite{Krasnov:1970bpz}.
    
It is now clear that we can obtain the action (\ref{action}) from (\ref{action-lambda}) by a contraction procedure that sets to zero the $AA$ terms in (\ref{action-lambda}). After this the curvature 2-forms $F^i$ are exact forms $F^i=dA^i\equiv da^i$, and the action coincides with (\ref{action}) restricted to the situation of "exact" 2-forms $\omega^i$. It is clear, however, that the action (\ref{action}) where $\omega^i$ do not need to be exact, is more general and makes sense also on closed 4-manifolds. 
\subsection{Relation between different formulations of (SD)GR. } 
All action discussed above have a direct relation to the action of General Relativity proposed by Plebanski \cite{Plebanski:1977zz}. 
This action reads
\begin{align}
    S[H,A,\Psi]&=\int  H^{i} F_{i} - \tfrac12 \left( \Psi_{ij} + \frac{\Lambda}{3} \delta_{ij}\right)  H^{i} H^{j}\,.
\end{align}
Here $H^i$ is a triple of 2-forms, and $A^i$ is an ${\rm SO}(3)$ gauge field. As before $\Psi^{ij}$ are "Lagrange multiplier" fields. The equation for $H^i$ is
\begin{align}\label{eqn-H}
    F_i= \left( \Psi_{ij} + \frac{\Lambda}{3} \delta_{ij}\right) H^j\,.
\end{align}

Plebanski action can be used to obtain an alternative description of GR in which only the fields $\Psi^{ij}$ and $A^i$ remain. This is done by solving the equation (\ref{eqn-H}) for the 2-form fields $H^i$. This gives
\begin{align}
    H^i&= \left( \Psi + \frac{\Lambda}{3} \mathtt{1}\right)^{-1}_{ij}  F^j\,.
\end{align}
Substituting this into the original action gives
\begin{align}
    S&= \tfrac12 \int F^i F^j (\Psi+ \Lambda \mathtt{1} )^{-1}_{ij} \,.
\end{align}
This action is still equivalent to the full Einstein-Hilbert action with a non-zero cosmological constant. The truncation to SDGR corresponds to an expansion of the matrix $(\Psi+ \Lambda \mathtt{1} )^{-1}$ in powers of $\Psi$, and then dropping all terms apart from 
\begin{align}
    S&= \tfrac12 \int F^i F^j \Psi_{ij} \,.
\end{align}
If one further drops the $AA$ term in the curvature one obtains the "flat" SDGR action (\ref{action}) that is the subject of this paper.

There is another action for flat SDGR, which is due to Siegel \cite{Siegel:1992wd}. It is obtained by dropping the cosmological constant term, as well as the $AA$ term from the curvature. In order to distinguish the new field, which is no longer a connection, from $A^i$, we change the name to $a^i$. The action is then
\begin{align}\label{action-flat-S1}
    S[H,A]&=\int H^i d a_i - \frac{1}{2}\Psi_{ij} H^i H^j\,.
\end{align}
The action in \cite{Siegel:1992wd} is a different, but a closely related one. To see this, we first note that the purpose of the last term containing $\Psi^{ij}$ is to impose the constraint that implies that $H^i$ is the self-dual part of the 2-form constructed from the frame. Thus, (\ref{action-flat-S1}) is equivalent to an action for the frame field and a triple of 1-forms 
\be\label{action-flat-S2}
S[e,A] = \int (e\wedge e)^i_{SD} da_i\,.
\ee
Here $(e\wedge e)^i_{SD}$ is the triple of self-dual 2-forms constructed from the frame $e$. As is explained in \cite{AbouZeid:2005dg}, the variation of this action with respect to $a^i$ gives an equation that implies that the self-dual part of the spin connection vanishes, which thus gives the correct description of gravitational instantons. 

The difference between (\ref{action-flat-S1}), (\ref{action-flat-S2}) and (\ref{action}) is in a different field content that is used to obtain the equation that guarantees that the self-dual part of the spin connection vanishes. In (\ref{action-flat-S1}), (\ref{action-flat-S2}) we have a metric-like field $H^i$ or a frame $e$, and the main equation that arises is a first order partial differential equation on this metric-like field. The action (\ref{action}) gives a "connection" description of gravitational instantons, in which the metric is constructed from the derivative $\omega^i=da^i$ of the connection-like field $a^i$. The main arising equation in this formalism is the algebraic equation (\ref{simplicity}) for the objects $\omega^i=da^i$.

The advantage of the formulation (\ref{action}) as compared to (\ref{action-flat-S2}) is that the covariant form of gauge-fixing that is necessary to do covariant perturbative calculations is much simpler for (\ref{action}) than for (\ref{action-flat-S2}). For the latter, the gauge-fixing is quite non-trivial and was recently described in \cite{Krasnov:2020bqr}. This reference contains the description relevant for full GR. The case of SDGR is obtained by a truncation. Nevertheless, even in the truncated case the gauge-fixing remains quite involved. In contrast, there is a much simpler covariant gauge-fixing that is available for the action (\ref{action}). This will be described below. This, together with the strong similarity to the known SDYM action (\ref{intro-sdym}) that we already discussed in the Introduction suggests that the action (\ref{action}) is more useful for explicit calculations than (\ref{action-flat-S2}).

\section{Perturbative expansion, amplitudes, light-cone}
\label{sec:perturb}
We now consider the action (\ref{action}) in the form (\ref{action-intro}) that is appropriate for expansion around the flat space background. The gauge symmetries of this action are diffeomorphisms as well as shifts of the 1-form potentials $a^i$ by exact 1-forms. Together, these can be described as follows
\begin{align}\label{gauge}
    \delta a^i&= d\xi^i + i_\eta H^i +i_\eta da^i\,, &
    \delta \Psi^{ij}&=i_\eta d \Psi^{ij}\,,
\end{align}
where we used $\mathcal{L}_\eta =\{d, i_\eta\}$ and absorbed $d (i_\eta\omega^i)$ into $\xi^i$. This gives the most convenient representation of the gauge symmetries, because in this form there are no derivatives of the vector fields $\eta^\mu$. The diffeomorphisms then act purely algebraically on the fields, which greatly simplifies the gauge-fixing, see below.
\subsection{Free fields and the spinor notation}
We have already quoted the perturbative expansion of the action (\ref{action}) around the flat space background in (\ref{action-pert-intro}). The background corresponds to the choice of the basic 2-forms $\omega^i$ being equal to the 't Hooft symbols $H^i$. 
To understand the structure of the first, kinetic term of the action (\ref{action-pert-intro}) it is very useful to pass to the spinor notation. 

The spinor translation of the fields $\Psi^{ij}, a^i$ is as follows. The object $a^i_\mu$ becomes $a^{AB}_\mu$, which is a one-form with values in symmetric rank two spinors. Indices $\mu,\nu,...=0,...,3$ are world indices; $A,B,C,...=1,2$ and $A',B',...=1,2$ are the spinor indices of the Lorentz algebra $sl(2,\mathbb{C})$. Translating into spinor indices also the spacetime (1-form) index $\mu$ we get an object $a^{AB}{}_{CC'}$, where now we have two different types of spinor indices. The tracefree field $\Psi^{ij}$ translates into the totally symmetric rank 4 spinor $\Psi^{ABCD}$. The spinor notation for the partial derivative is $\partial_{AA'}$. The exterior derivative $da^i$ becomes the following spinorial object $\partial_{CC'} a^{AB}{}_{DD'}$. The first, kinetic term in  (\ref{action-pert-intro}) involves taking the wedge product of this with self-dual 2-forms $H^i$. This causes the self-dual projection of the object $\partial_{CC'} a^{AB}{}_{DD'}$, because self-dual forms only pair non-trivially with self-dual one under the wedge product. The self-dual projection of the 2-form $\partial_{CC'} a^{AB}{}_{DD'}$ is the object $\partial_{CC'} a^{AB}{}_{D}{}^{C'}$ in which the primed spinor index is contracted. All in all, the spinor translation of the kinetic term in  (\ref{action-pert-intro})  reads
\be
S^{(2)}[\Psi,a] = \int \Psi_{ABCD} \partial^{A}{}_{A'} a^{BCDA'}\,.
\ee
We have rescaled the field $\Psi^{ABCD}$ to absorb the dimensionful quantity $M^2$ into it, so as to get the canonically normalised kinetic term. This means that the potential field $a$ has mass dimension one, while $\Psi$ has mass dimension two. 

We now introduce a very convenient notation. To save indices, from now on, the indices that belong to a group of symmetric (or to be symmetrized) indices can be denoted by the same letter. The kinetic term written in this notation reads
\begin{align}\label{action-diff-gf}
    S&=\int \Psi_{AAAA} \partial^{A}{}_{A'} a^{AAAA'}\,.
\end{align}
The corresponding equations of motion are
\begin{align}\label{feqs-lin}
    \pl\fdu{B}{A'}\Psi^{AAAB}&=0\,, && \pl\fud{A}{B'}a^{AAA,B'}=0\,.
\end{align}
These equations are well-known \cite{Penrose:1965am,Hughston:1979tq,Eastwood:1981jy,Woodhouse:1985id}. They describe the helicity $+2$ and $-2$ states of the free graviton. The first one is just a part of the Bianchi identities for the Weyl tensor. As expected, we have the correct free limit where the theory describes free massless spin-two degrees of freedom.

\subsection{Gauge-fixing} 
The gauge symmetries (\ref{gauge}) linearized around the flat background are
\begin{align}
    \delta a^{AB}&= d\xi^{AB}+ \eta\fud{A}{B'}e^{BB'} &
    \delta \Psi^{AAAA}&=0\,.
\end{align}
Here $e^{AA'}$ is the background vierbein 1-form, and the last piece in the gauge transformation for the potential fields $a^i$ originates from $i_\eta H^{AB}$, where we converted vector-field $\eta^\mu$ into a bi-spinor $\eta^{AA'}=e^{AA'}_\mu \eta^\mu$
with the help of background vierbein $e^{AA'}_\mu$. Now, $a^{AB}$ can be decomposed into two irreducible spin-tensors:
\begin{align}
    a^{AB}&= e_{CC'} \Phi^{ABC,C'} + e\fud{A}{B'} \Phi^{BB'}
\end{align}
of type $(3,1)$ and $(1,1)$, respectively. It is clear that $\eta$-symmetry allows us to gauge away the second component in a peaceful algebraic way. As the result we have a theory of two irreducible spin-tensors, $\Psi^{ABCD}$ and $\Phi^{ABC,A'}$, as the physical fields and with the following linearised gauge transformations
\begin{align}
    \delta \Phi^{AAA,A'}&= \pl^{AA'}\xi^{AA}\,, & \delta \Psi^{AAAA}&=0\,,
\end{align}
where we used our convention that a set of spinor indices that is symmetrised is denoted by the same letter. 

The linearised action (\ref{action-diff-gf}) does not depend on the $\Phi^{BB'}$ component of the connection, and in this sense already has the diffeomorphism symmetry gauge-fixed. There is still the symmetry $\delta \Phi^{AAA,A'}= \pl^{AA'}\xi^{AA} $ that this action is invariant under. To gauge-fix it, and produce a kinetic term that can be inverted to obtain the propagator, we use a variant of the Lorentz gauge. The same gauge-fixing procedure but for the SDGR with non-zero cosmological constant has been described in \cite{Krasnov:2016emc}. 

The idea is to add a Lagrange multiplier term imposing the Lorentz gauge condition
\be
\pl^{BB'}\Phi_{AAB,B'}=0\,.
\ee
This is added with a Lagrange multiplier $\Psi^{AA}$. One then notices that the field $\Psi_{AAAA}$ already present in the action can be combined with the new Lagrange multiplier field $\Psi^{AA}$ to produce the new field
\be\label{gauge-fixing}
\tilde{\Psi}^{ABCD} := \Psi^{ABCD} + \Psi^{(AB}\epsilon^{C)D} \,.
\ee
The new field is totally symmetric in its first 3 spinor indices, and consists of two different irreducible components $\Psi^{ABCD}$ and $\Psi^{AB}$. Returning to the convention of a repeated spinor index to denote a group of spinor indices that is totally symmetric, the gauge-fixed action becomes 
\begin{align}\label{action-diff-gfB}
    \int \tilde{\Psi}^{AAAB} \pl\fdu{B}{B'} \Phi_{AAA,B'} \,.
\end{align}
This action depends on two fields $\Phi^{AAA,B'}$ and $\tilde{\Psi}^{AAAB}$ that both contain the same number of components. The operator that maps one into the other is a version of the (chiral) Dirac operator, and is non-degenerate. Its inverse is the propagator of the theory. 

\subsection{Tree-level amplitudes}
\label{sec:trees}

\paragraph{Polarisation spinors}

We start by discussing the momentum space representation of the solutions of the linearised field equations (\ref{feqs-lin}). It is clear that these equations describe the two helicities of the graviton asymmetrically. One of the two helicities resides in the "potential" field $\Phi^{AAA,B'}$, while the other helicity resides in $\Psi^{AAAA}$. Let us agree that it is the negative helicity that is described by the connection. The corresponding polarisation tensor is then
\be
\epsilon^-_{AAA,A'}(k) = M \frac{q_A q_A q_A k_{A'}}{ (qk)^{3}}\,.
\ee
Here we have a null momentum $k_{AA'}=k_{A} k_{A'}$, $q_A$ is an auxiliary spinor, and $(qk):= q^A k_A$ is the spinor contraction. We have also included a dimensionful parameter $M$ to an appropriate power in front, so that the polarisation spinor is dimensionless.
The polarisation spinor introduced satisfies the first of the equations in (\ref{feqs-lin}).

The polarisation spinor describing the opposite, positive helicity is an object with only unprimed spinor indices and is given by
\be\label{negative}
\epsilon^+_{AAAA} = M^{-1} k_{A} k_A k_A k_A \,.
\ee
The mass parameter in front is so that the mass dimension of this spinor is one. This polarisation spinor satisfies the momentum space version of the second of the equations in (\ref{feqs-lin}). 

\paragraph{Amplitude characterisation of the cubic vertex}

We now evaluate the cubic vertex in (\ref{action-pert-intro}) on shell, by inserting into it appropriate polarisation spinors. Given that we have absorbed the mass parameter $M^2$ into $\Psi$ to give it the mass dimension two, the interaction takes the form
\be
\frac{1}{2M^2} \int \Psi^{AAAA} da_{AA} \wedge da_{AA}\,.
\ee

On negative helicity states $\epsilon^-(k)$ the 2-forms $d a_{AA}$ have only the ASD parts, because their SD parts vanish in view of the equation satisfied by these states. So, the only non-vanishing part of $d a_{AA}$ is the spinor
\be
M \frac{q_A q_A}{ (qk)^2} k_{A'} k_{B'}\,.
\ee
Note that, apart from the dimensionful prefactor, this is the usual polarisation state for a single negative helicity graviton.

The cubic interaction in (\ref{action-pert-intro}) contains the wedge product of two such $da$ factors. This converts into the contraction of the primed indices, which gives
\be
M^2 [12]^2 \frac{q_A q_A q_A q_A}{ (q1)^{2}(q2)^{2}}\,.
\ee
We now contract this with a positive helicity polarisation spinor (\ref{negative}), and multiply the result by $M^{-2}$ present in front of the action. The amplitude is then given by
\be
{\cal A}^{--+} = \frac{1}{M} [12]^2 \frac{(q3)^{4}}{ (q1)^{2}(q2)^{2}}\,.
\ee
Eliminating factors of the auxiliary spinor $q$ using the momentum conservation identities $(q3)/(q1)=-[12]/[32], (q3)/(q2)=-[21]/[31]$ we get
\be\label{amplitude}
{\cal A}^{--+} = \frac{1}{M} \frac{[12]^6}{ [23]^{2}[13]^{2}} \,.
\ee
This is the usual result in gravity. This also allows us to identify $M=M_{Pl}$.

\subsection{Light-cone gauge }

Another very useful characterisation of the theory can be obtained by passing into the light-cone gauge. This reduces everything to the physical degrees of freedom only. Light-cone gauge also allows to perform explicit computations of Feynman diagrams since SDGR turns out to be a rather simple theory of two 'scalar' fields representing helicity $\pm2$ states.

As the first step we impose the light-cone gauge by setting $\Phi^{AB+,+'}=0$.\footnote{We change the range of indices from $1,2$ to  $A=+,-$, $A'=+',-'$, etc.} Then, the physical degree of freedom is in $\Phi_{-2}=(\pl^{++'})^{-2}\Phi^{---,+'}$ and $\Phi^{---,-'}=\pl^{+-'}(\pl^{++'})^{-1}\Phi^{---,+'}$ is an auxiliary field. As for $\Psi_{ABCD}$, the physical degree of freedom resides in $\Phi_{+2}=(\pl^{++'})^2\Psi_{----}$, the rest being auxiliary fields. Plugging this into the full action we end up with the Siegel action \cite{Siegel:1992wd}
\begin{align}\label{lightcone}
    S&= \int (\Phi_{-2}\square \Phi_{+2} + M_{Pl}\,\epsilon_{A'B'}\epsilon_{C'D'} \Phi_{+2}\pl^{+A'}\pl^{+C'}\Phi_{-2}\pl^{+B'}\pl^{+D'}\Phi_{-2})\,,
\end{align}
where we also introduced a coupling constant $g$. 
We note that $\Phi_{\pm2}$ are related via $\phi_{\pm s}=(\pl^{++'})^{\mp 2}\Phi_{\pm s }$ to the fields $\phi_{\pm s}$ that transform canonically under the Lorentz transformations, see e.g. \cite{Bengtsson:1986kh,Metsaev:1991mt} and \cite{Siegel:1992wd} for more detail on light-cone manipulations.

\section{Gauge algebra of SDGR} 
\label{sec:algebra}
It is also interesting to discuss the gauge algebra of SDGR in flat and anti-de Sitter space. Let us start with the Lorentz $L_{AA}$, $L_{A'A'}$ and translations $P_{AA'}$ generators of the anti-de Sitter algebra $so(3,2)\sim sp(4,\mathbb{R})$:
\besubeqs\label{adsrelations} 
\begin{align}
    [L_{AB},P_{CC'}]&= \epsilon_{BC}P_{AC'}+\epsilon_{AC}P_{BC'}\,,\\
    [L_{A'B'},P_{CC'}]&= \epsilon_{B'C'}P_{CA'}+\epsilon_{A'C'}P_{CB'}\,,\\
    [P_{AA'},P_{BB'}]&= \epsilon_{A'B'} L_{AB}+\epsilon_{AB} L_{A'B'}\,,\\
    [L_{AB},L_{CD}]&= \epsilon_{BC} L_{AD}+\text{3 more}\,,\\
    [L_{A'B'},L_{C'D'}]&= \epsilon_{B'C'} L_{A'D'}+\text{3 more}\,.
\end{align}
\esubeqs
Introducing a gauge field of this algebra, $A=\tfrac12 \omega^{AA}L_{AA}+\tfrac12 \omega^{A'A'}L_{A'A'}+e^{AA'}P_{AA'}$ the curvature $F(A)$ of $A$ decomposes into
\besubeqs
\begin{align}
    d \omega^{AA}+\omega\fud{A}{C}\wedge\omega^{CB}+ e\fud{A}{B'}\wedge e^{AB'}&=R^{AA}\,,\\
    d e^{AA'}+\omega\fud{A'}{B'}\wedge e^{AB'}+\omega\fud{A}{B}\wedge e^{BA'} &=T^{AA'}\,,\\
    d \omega^{A'A'}+\omega\fud{A'}{C'}\wedge\omega^{C'B'}+ e\fdu{B}{A'}\wedge e^{BA'}&=R^{A'A'}\,,
\end{align}
\esubeqs
the second one being torsion and the other two being two components of the curvature two-form. The first curvature is the one that is used in the AdS-SDGR with $R^{AB}=F^{AB}+e\fud{A}{B'}\wedge e^{AB'}$. The $\Psi$-equations of motion guarantee that $F^{AA}$ can be expressed as $e\fud{A}{B'}\wedge e^{AB'}$. 

In order to reproduce the structures relevant for the flat space SDGR, we can take a limit where $L_{AA}$ becomes central and $L_{A'A'}$ disappears from $[P,P]$, i.e.
\besubeqs\label{weirdlimit}
\begin{align}
    [L_{AB},P_{CC'}]&= 0\,,\\
    [L_{A'B'},P_{CC'}]&= \epsilon_{B'C'}P_{CA'}+\epsilon_{A'C'}P_{CB'}\,,\\
    [P_{AA'},P_{BB'}]&= \epsilon_{A'B'} L_{AB}\,,\\
    [L_{AB},L_{CD}]&= 0\,,\\
    [L_{A'B'},L_{C'D'}]&= \epsilon_{B'C'} L_{A'D'}+\text{3 more}\,.
\end{align}
\esubeqs
Note that this limit is rather different from the flat space limit where the only change as compared to \eqref{adsrelations} is 
\begin{align}
    [P_{AA'},P_{BB'}]&= 0\,.
\end{align}
The limiting algebra \eqref{weirdlimit} takes advantage of the specific structure that is available only in $4d$: there are chiral $L_{AA}$ and anti-chiral $L_{A'A'}$ generators and we can treat them differently. The curvatures corresponding to \eqref{weirdlimit}  are
\besubeqs
\begin{align}
    d \omega^{AA}+ e\fud{A}{B'}\wedge e^{AB'}&=P^{AA}\,,\\
    d e^{AA'}+\omega\fud{A'}{B'}\wedge e^{AB'} &=T^{AA'}\,,\\
    d \omega^{A'A'}+\omega\fud{A'}{C'}\wedge\omega^{C'B'}&=P^{A'A'}\,.
\end{align}
\esubeqs
Our action takes advantage of $P^{AA}=F^{AA}+e\fud{A}{B'}\wedge e^{AB'}$ only. 

Let us discuss the relation to the well-known oscillator realisation of $so(5)\sim sp(4)$. One takes four operators with canonical commutation relations, $[y_A,y_B]=i\epsilon_{AB}$, $[\bry_{A'},\bry_{B'}]=i\epsilon_{A'B'}$ and defines 
\begin{align}
    L_{AB}&=\tfrac{-i}{2}\{y_A,y_B\}\,, &
    L_{A'B'}&=\tfrac{-i}{2}\{\bry_{A'},\bry_{B'}\}\,, &
    P_{AA'}&=\tfrac{-i}{2}\{y_A,\bry_{A'}\}\,.
\end{align}
These generators obey \eqref{adsrelations}. The limiting algebra \eqref{weirdlimit} can be obtained in the commutative limit $[y_A,y_B]=0$, while leaving $[\bry_{A'},\bry_{B'}]=i\epsilon_{A'B'}$ untouched. 

The construction above has an obvious higher spin generalization. In the higher spin case the gauge algebra is the even subalgebra of the Weyl algebra $A_2$ \cite{Vasiliev:1986qx}, i.e. its elements are even functions $f(y,\bry)=f(-y,-\bry)$. Similarly to SDGR, the limiting algebra is the commutative limit in $y_A$. Despite being commutative the algebra still features a certain deformation to $A_\infty$- and $L_\infty$-algebras described in \cite{Sharapov:2018kjz} (the deformation is due to a $\mathbb{Z}_2$-orbifold: Poisson manifolds with discrete symmetries can have more deformations in the sense of deformation quantization provided one considers the corresponding Poisson orbifold). 

\section{Discussion}

In this paper we have given a new simple action formulation of self-dual gravity that is appropriate for describing gravitational instantons with zero scalar curvature. Many elements of this description can be found in the mathematical literature \cite{Donaldson2006TwoformsOF},\cite{FINE_2017}, but the action (\ref{action}) appears to be new.

We have seen that the action (\ref{action}), when expanded around an appropriate background $\omega^i=H^i$, gives rise to the kinetic term (\ref{action-diff-gf}). The corresponding linearised field equations describe two different graviton helicities, one contained in the field $\Phi^{AAA,B'}$ and the other in $\Psi^{AAAA}$. We have also seen that the arising interaction, which is of the form $\Psi da da$, correctly reproduces the 3 point graviton scattering amplitude (\ref{amplitude}). The action (\ref{action}) also gives rise to the familiar pattern (\ref{lightcone}) in the light-cone gauge. 

One of the most intriguing aspects of the new formulation of SDGR is that the structure of the arising kinetic term (\ref{action-diff-gf}) literally mimics the structure familiar from the SDYM case. Indeed, as discussed in details in \cite{Krasnov:2016emc}, the spinor translation of the first, kinetic term in (\ref{action-SDYM}) is given by $\Psi^{AB} \partial_{B}{}^{B'} A_{AB'}$. We thus see that the change in the case of gravity is to add two more unprimed spinor indices. The gauge-fixing that is most useful in the case of SDYM, see \cite{Krasnov:2016emc}, is also the complete analog of the one in (\ref{gauge-fixing}). Thus, the kinetic terms of SDGR and SDYM can be treated in exact parallel, and the arising propagator is the inverse of an appropriate chiral Dirac operator. 

It is interesting that this suggests a new perspective on the colour/kinematics duality and the double copy structure \cite{Bern:2010ue} of gravity. Indeed, the double copy prescription is to first write the YM amplitudes so that the colour/kinematic symmetry is manifest. The second step is to replace the "colour" numerators with the "kinematic" ones. It is interesting that one can pass from the kinetic term of SDYM in (\ref{action-SDYM}) to the kinetic term of SDGR in (\ref{action-pert-intro}) simply by replacing the Lie algebra "colour" index of $\Psi^{ia}, A^a$ with the "chiral" index $i=1,2,3$ that enumerates the generators of the self-dual chiral half of the Lorentz Lie algebra. This suggests that there is a link between the chiral half of the Lorentz algebra and the mysterious "kinematics" Lie algebra whose existence is suggested by the statement of the colour/kinematics duality, see \cite{Fu:2016plh} for an attempt at identification of this Lie algebra. It would be very interesting to see if the new formulation of SDGR can shed any light on the double copy structure of gravity, even if only in the self-dual sector, see \cite{Campiglia:2021srh} for some recent work on the double copy in the self-dual sector. We hope to return to this question in another publication. 

\section*{Acknowledgments}
\label{sec:Aknowledgements}
We are grateful to Alexey Sharapov for useful discussions. KK is grateful to Joel Fine for answering questions on the description of hyper-K\"ahler manifolds that features in this paper. The work of E.S. was supported by the Russian Science Foundation grant 18-72-10123 in association with the Lebedev Physical Institute.

\footnotesize
\providecommand{\href}[2]{#2}\begingroup\raggedright\endgroup


\begin{thebibliography}{10}

\bibitem{Krasnov:2016emc}
K.~Krasnov, ``{Self-Dual Gravity},''
  \href{http://dx.doi.org/10.1088/1361-6382/aa65e5}{{\em Class. Quant. Grav.}
  {\bfseries 34} no.~9, (2017) 095001},
  \href{http://arxiv.org/abs/1610.01457}{{\ttfamily arXiv:1610.01457
  [hep-th]}}.

\bibitem{Chalmers:1996rq}
G.~Chalmers and W.~Siegel, ``{The Selfdual sector of QCD amplitudes},''
  \href{http://dx.doi.org/10.1103/PhysRevD.54.7628}{{\em Phys. Rev.} {\bfseries
  D54} (1996) 7628--7633},
\href{http://arxiv.org/abs/hep-th/9606061}{{\ttfamily arXiv:hep-th/9606061
  [hep-th]}}.

\bibitem{Bittleston:2020hfv}
R.~Bittleston and D.~Skinner, ``{Twistors, the ASD Yang-Mills equations, and 4d
  Chern-Simons theory},'' \href{http://arxiv.org/abs/2011.04638}{{\ttfamily
  arXiv:2011.04638 [hep-th]}}.

\bibitem{Siegel:1992wd}
W.~Siegel, ``{Selfdual N=8 supergravity as closed N=2 (N=4) strings},''
  \href{http://dx.doi.org/10.1103/PhysRevD.47.2504}{{\em Phys. Rev. D}
  {\bfseries 47} (1993) 2504--2511},
  \href{http://arxiv.org/abs/hep-th/9207043}{{\ttfamily arXiv:hep-th/9207043}}.

\bibitem{AbouZeid:2005dg}
M.~Abou-Zeid and C.~M. Hull, ``{A Chiral perturbation expansion for gravity},''
  \href{http://dx.doi.org/10.1088/1126-6708/2006/02/057}{{\em JHEP} {\bfseries
  02} (2006) 057}, \href{http://arxiv.org/abs/hep-th/0511189}{{\ttfamily
  arXiv:hep-th/0511189}}.

\bibitem{Bern:2010ue}
Z.~Bern, J.~J.~M. Carrasco, and H.~Johansson, ``{Perturbative Quantum Gravity
  as a Double Copy of Gauge Theory},''
  \href{http://dx.doi.org/10.1103/PhysRevLett.105.061602}{{\em Phys. Rev.
  Lett.} {\bfseries 105} (2010) 061602},
\href{http://arxiv.org/abs/1004.0476}{{\ttfamily arXiv:1004.0476 [hep-th]}}.

\bibitem{Metsaev:1991mt}
R.~R. Metsaev, ``{Poincare invariant dynamics of massless higher spins: Fourth
  order analysis on mass shell},''
{\em Mod. Phys. Lett.} {\bfseries A6} (1991) 359--367.

\bibitem{Metsaev:1991nb}
R.~R. Metsaev, ``{$S$ matrix approach to massless higher spins theory. 2: The
  Case of internal symmetry},''
{\em Mod. Phys. Lett.} {\bfseries A6} (1991) 2411--2421.

\bibitem{Ponomarev:2016lrm}
D.~Ponomarev and E.~D. Skvortsov, ``{Light-Front Higher-Spin Theories in Flat
  Space},'' {\em J. Phys.} {\bfseries A50} no.~9, (2017) 095401,
\href{http://arxiv.org/abs/1609.04655}{{\ttfamily arXiv:1609.04655 [hep-th]}}.

\bibitem{Skvortsov:2018jea}
E.~D. Skvortsov, T.~Tran, and M.~Tsulaia, ``{Quantum Chiral Higher Spin
  Gravity},'' {\em Phys. Rev. Lett.} {\bfseries 121} no.~3, (2018) 031601,
\href{http://arxiv.org/abs/1805.00048}{{\ttfamily arXiv:1805.00048 [hep-th]}}.

\bibitem{Skvortsov:2020wtf}
E.~Skvortsov, T.~Tran, and M.~Tsulaia, ``{More on Quantum Chiral Higher Spin
  Gravity},'' {\em Phys. Rev.} {\bfseries D101} no.~10, (2020) 106001,
\href{http://arxiv.org/abs/2002.08487}{{\ttfamily arXiv:2002.08487 [hep-th]}}.

\bibitem{Skvortsov:2020gpn}
E.~Skvortsov and T.~Tran, ``{One-loop Finiteness of Chiral Higher Spin
  Gravity},''
\href{http://arxiv.org/abs/2004.10797}{{\ttfamily arXiv:2004.10797 [hep-th]}}.

\bibitem{Skvortsov:2018uru}
E.~Skvortsov, ``{Light-Front Bootstrap for Chern-Simons Matter Theories},''
  {\em JHEP} {\bfseries 06} (2019) 058,
\href{http://arxiv.org/abs/1811.12333}{{\ttfamily arXiv:1811.12333 [hep-th]}}.

\bibitem{Krasnov:2021nsq}
K.~Krasnov, E.~Skvortsov, and T.~Tran, ``{Actions for Self-dual Higher Spin
  Gravities},''
\href{http://arxiv.org/abs/2105.12782}{{\ttfamily arXiv:2105.12782 [hep-th]}}.

\bibitem{Donaldson2006TwoformsOF}
S.~Donaldson, ``Two-forms on four-manifolds and elliptic equations,'' {\em
  arXiv: Differential Geometry} (2006) .

\bibitem{FINE_2017}
J.~Fine, J.~D. Lotay, and M.~Singer, ``The space of hyperk\"{a}hler metrics on
  a 4-manifold with boundary,''
  \href{http://dx.doi.org/10.1017/fms.2017.3}{{\em Forum of Mathematics, Sigma}
  {\bfseries 5} (2017) }.

\bibitem{Krasnov:1970bpz}
K.~Krasnov, \href{http://dx.doi.org/10.1017/9781108674652}{{\em {Formulations
  of General Relativity}}}.
\newblock Cambridge Monographs on Mathematical Physics. Cambridge University
  Press, 11, 2020.

\bibitem{Hitchin:2001rw}
N.~J. Hitchin, ``{Stable forms and special metrics},''
  \href{http://arxiv.org/abs/math/0107101}{{\ttfamily arXiv:math/0107101}}.

\bibitem{Plebanski:1977zz}
J.~F. Plebanski, ``{On the separation of Einsteinian substructures},''
  \href{http://dx.doi.org/10.1063/1.523215}{{\em J. Math. Phys.} {\bfseries 18}
  (1977) 2511--2520}.

\bibitem{Krasnov:2020bqr}
K.~Krasnov and Y.~Shtanov, ``{Chiral perturbation theory for GR},''
  \href{http://dx.doi.org/10.1007/JHEP09(2020)017}{{\em JHEP} {\bfseries 09}
  (2020) 017}, \href{http://arxiv.org/abs/2007.00995}{{\ttfamily
  arXiv:2007.00995 [hep-th]}}.

\bibitem{Penrose:1965am}
R.~Penrose, ``{Zero rest mass fields including gravitation: Asymptotic
  behavior},''
\href{http://dx.doi.org/10.1098/rspa.1965.0058}{{\em Proc. Roy. Soc. Lond.}
  {\bfseries A284} (1965) 159}.

\bibitem{Hughston:1979tq}
L.~P. Hughston, R.~S. Ward, M.~G. Eastwood, M.~L. Ginsberg, A.~P. Hodges, S.~A.
  Huggett, T.~R. Hurd, R.~O. Jozsa, R.~Penrose, A.~Popovich, {\em et~al.},
  eds., {\em {Advances in twistor theory}}.
\newblock
1979.
\newblock

\bibitem{Eastwood:1981jy}
M.~G. Eastwood, R.~Penrose, and R.~O. Wells, ``{Cohomology and Massless
  Fields},''
\href{http://dx.doi.org/10.1007/BF01942327}{{\em Commun. Math. Phys.}
  {\bfseries 78} (1981) 305--351}.

\bibitem{Woodhouse:1985id}
N.~M.~J. Woodhouse, ``{Real methods in twistor theory},''
\href{http://dx.doi.org/10.1088/0264-9381/2/3/006}{{\em Class. Quant. Grav.}
  {\bfseries 2} (1985) 257--291}.

\bibitem{Bengtsson:1986kh}
A.~K.~H. Bengtsson, I.~Bengtsson, and N.~Linden, ``{Interacting Higher Spin
  Gauge Fields on the Light Front},''
{\em Class. Quant. Grav.} {\bfseries 4} (1987) 1333.

\bibitem{Vasiliev:1986qx}
M.~A. Vasiliev, ``Extended higher spin superalgebras and their realizations in
  terms of quantum operators,''
{\em Fortsch. Phys.} {\bfseries 36} (1988) 33--62.

\bibitem{Sharapov:2018kjz}
A.~Sharapov and E.~Skvortsov, ``{$A_\infty$ algebras from slightly broken
  higher spin symmetries},''
  \href{http://dx.doi.org/10.1007/JHEP09(2019)024}{{\em JHEP} {\bfseries 09}
  (2019) 024},
\href{http://arxiv.org/abs/1809.10027}{{\ttfamily arXiv:1809.10027 [hep-th]}}.

\bibitem{Fu:2016plh}
C.-H. Fu and K.~Krasnov, ``{Colour-Kinematics duality and the Drinfeld double
  of the Lie algebra of diffeomorphisms},''
  \href{http://dx.doi.org/10.1007/JHEP01(2017)075}{{\em JHEP} {\bfseries 01}
  (2017) 075}, \href{http://arxiv.org/abs/1603.02033}{{\ttfamily
  arXiv:1603.02033 [hep-th]}}.
  
\bibitem{Campiglia:2021srh}
M.~Campiglia and S.~Nagy,
``A double copy for asymptotic symmetries in the self-dual sector,''
JHEP \textbf{03}, 262 (2021)
doi:10.1007/JHEP03(2021)262
[arXiv:2102.01680 [hep-th]].

\end{thebibliography}
\end{document}